\documentclass[11pt]{article}

\usepackage{latexsym}
\usepackage{epsfig}

  \large
\begin{document}
\thispagestyle{empty}
\newcommand{\R}{{\mathchoice{\hbox{$\sf\textstyle I\hspace{-.15em}R$}}
{\hbox{$\sf\textstyle I\hspace{-.15em}R$}} {\hbox{$\sf\scriptstyle
I\hspace{-.10em}R$}} {\hbox{$\sf\scriptscriptstyle
I\hspace{-.11em}R$}}}}
\newcommand{\Dirac}{\rlap{\hspace{-.5mm} \slash} D}
\renewcommand{\theequation}{\arabic{section}.\arabic{equation}}
\renewcommand{\d}{\displaystyle}
\newcommand{\be}{\begin{eqnarray}}
\newcommand{\ee}{\end{eqnarray}}
\newcommand{\mat}{\left ( \begin{array}{cc}}
\newcommand{\emat}{\end{array} \right )}
\newcommand{\matf}{\left ( \begin{array}{cccc}}
\newcommand{\ematf}{\end{array} \right )}
\hfill{SUNY-NTG-00/61}

\vspace{2cm}
\begin{center}
{\Large\bf Statistical properties of the spectrum of \\[0.4cm]
the QCD Dirac operator at low energy} \vspace{8mm}

D. Toublan$^{1,2}$ and J.J.M. Verbaarschot$^2$ \vskip 0.2cm {\it
$^1$Loomis Laboratory of Physics, University of Illinois,
Urbana-Champaign, IL 61801, USA
\\
$^2$Department of Physics and Astronomy, State University of New
York, Stony Brook, NY 11794, USA} \vskip 1.5cm

{\bf Abstract}
\end{center}
{\footnotesize We analyze the statistical properties of the
spectrum of the QCD Dirac operator at low energy in a finite box
of volume $L^4$ by means of partially quenched Chiral Perturbation
Theory, a low-energy effective field theory based on the 
symmetries of QCD. We derive the two-point spectral correlation
function from the discontinuity of the chiral susceptibility. For
eigenvalues much smaller than $m_c=F^2/\Sigma L^2$, where $F$ is
the pion decay constant and $\Sigma$ is the absolute value of the
quark condensate, our result for the two-point correlation
function coincides with the result previously obtained from chiral
Random Matrix Theory (chRMT). The departure from the chRMT result
above that scale is due to the contribution of the nonzero  
momentum modes. In terms of the variance of the number of
eigenvalues in an interval containing $n$ eigenvalues on average,
it amounts to  a crossover from a $\log n$-behavior to a   
$n^2 \log n$-behavior.}
 

\vfill\eject

\pagestyle{plain}

\noindent
\section{Introduction}

Since the seminal work by Wigner, Dyson and Mehta
\cite{Wign51,Dyso62,Mehta} on the Random Matrix Theory description
of level correlations in nuclei, the problem of level statistics
has been analyzed in great detail for many different quantum
systems (see \cite{HDgang} for a recent review). In QCD, a
statistical analysis can be applied to the eigenvalues of the
Dirac operator $i\Dirac$ defined by \be i\Dirac \phi_k =
i\lambda_k \phi_k. \ee The Euclidean QCD partition function for
$N_f$ quarks of mass $m_f$ is given by \be
  \label{ZQCD}
  Z^{\rm QCD}  &=& \int\! [dA]_{\nu}  \prod_{f=1}^{N_{f}}
\det(i\Dirac + m_f) ~e^{-S_{YM}[A]}, \ee and can thus be expressed
as \be
  Z^{\rm QCD}
&=& \langle \prod_{f=1}^{N_f} \prod_k (i \lambda_k+ m_f)
\rangle_{\rm YM}. \ee Here, $\langle \cdots \rangle_{\rm YM}$
denotes an average over the Yang-Mills partition function. This
shows that, although the eigenvalues cannot be observed directly,
their properties are of fundamental importance to the physics of
QCD.

For small enough energies, below the so-called Thouless energy
\cite{Vplb,james,OTV,DOTV},  the eigenvalues of the QCD Dirac operator 
are strongly
correlated, and their correlations are given by chiral Random
Matrix Theory (chRMT) \cite{SV,V,Maetal}. For energy differences much  
larger than the Thouless energy but much smaller than
$\Lambda_{\rm QCD}$, the eigenvalues of the QCD Dirac operator
show much weaker correlations that are different from chRMT. In
this domain, the eigenvalue correlations can be computed
perturbatively  by means of partially quenched Chiral Perturbation
Theory (pqChPT). This is a low-energy effective theory based only
on the symmetries of QCD formulated to probe the spectrum  of the
QCD Dirac operator. Finally, for energies beyond $\Lambda_{\rm
QCD}$ the eigenvalues are uncorrelated. 
 These different
domains have been identified in lattice QCD simulations
\cite{Gock,Berb98c,Berb99}, and can be derived from  
random hopping models with the chiral symmetry of the QCD
partition function
\cite{Gade,simons-altland,Takahashi,Guhr-Wilke-Weidi,Takanew,GWil}. 
The occurrence of Random Matrix behavior in                     
the Dirac spectrum                        
can be understood naturally from universality arguments.      

The Thouless energy can be studied quantitatively by means of pqChPT. 
>From the  analysis of the average spectral density of the              
QCD Dirac operator,  $\rho(\lambda)$, we found                        
\cite{OTV,DOTV,TV} that the Thouless energy  is                     
given by $m_c=F^2/\Sigma L^2$, where $F$ is the pion decay constant   
and $\Sigma$ is the                                                   
absolute value of the quark condensate. The $1/L^2$-dependence of     
$m_c$                                                                 
is well-known from the theory of mesoscopic systems     
(see for example \cite{Montambaux}). In addition to                   
$\Lambda_{\rm QCD}$ there is one                                      
other important energy scale in the Dirac spectrum:                   
the smallest nonzero eigenvalue $\lambda_{\rm min}$.                  
It is directly related to the spectral density near zero virtuality  
and is
therefore given by the Banks-Casher formula \cite{BC}:                
$\lambda_{\rm min}=1/\rho(0) =\pi/\Sigma V$. With the terminology
adopted from the study of disordered mesoscopic systems
\cite{Montambaux} we can  thus distinguish four different     
domains in the Dirac 
spectrum. The quantum domain and the ergodic domain are separated
by $\lambda_{\rm min}$, the ergodic domain and the diffusive
domain are separated by the Thouless energy $m_c$, and the   
diffusive domain and the ballistic domain are separated by
$\Lambda_{\rm QCD}$.

In this article, we turn our attention to correlations of  Dirac
eigenvalues, i.e. the fluctuations about the average behavior of
the spectrum. For this purpose we consider multi-level correlation
functions. We focus our study on the average connected two-point
spectral correlation function, $\rho_c(\lambda_1,\lambda_2)$,
defined by
\begin{eqnarray}
\label{rho2}
 \rho_c(\lambda_1,\lambda_2)&=&\langle \sum_{i,j}
\delta(\lambda_1-\lambda_i)
\delta(\lambda_2-\lambda_j)\rangle_{\rm QCD}
\nonumber \\
&&- \langle \sum_k  \delta(\lambda_1-\lambda_k) \rangle_{\rm QCD}
\; \langle \sum_k  \delta(\lambda_2-\lambda_k) \rangle_{\rm QCD}
\nonumber \\
&\equiv& \delta(\lambda_1 -\lambda_2) \langle \sum_k
\delta(\lambda_2-\lambda_k) \rangle_{\rm QCD}
+R(\lambda_1,\lambda_2) .
\end{eqnarray}
where $\langle \cdots \rangle_{\rm QCD}$ denotes the average over 
the QCD partition function, and the $\lambda_k$ are the eigenvalues of 
the Dirac operator. In the last line of this equation we have
decomposed the correlation function into a term containing the
self-correlations and the two-point cluster function,
$R(\lambda_1,\lambda_2)$. Both terms enter in the disconnected
scalar susceptibility which is a more natural object in a field
theory context. It is defined by
\begin{eqnarray}
\chi&=&\frac{1}{V} \; \partial_{m_1}
\partial_{m_2} \; {\rm ln} \, Z_{\rm QCD}  \nonumber \\
     &=& \frac1V \langle \sum_k (i \lambda_k+m_1)^{-1} \; \sum_j (i
     \lambda_j+m_2)^{-1}  \rangle_{\rm QCD} \\
     &&-\frac1V \langle \sum_k (i
     \lambda_k+m_1)^{-1} \rangle_{\rm QCD} \; \langle \sum_j (i
     \lambda_j+m_2)^{-1} \rangle_{\rm QCD} \nonumber \\
&=& \frac1V \int_{-\infty}^\infty d\lambda_1 \;
\int_{-\infty}^\infty d\lambda_2 \;
\frac{\rho_c(\lambda_1,\lambda_2)}{(i \lambda_1+m_1) (i
\lambda_2+m_2) } . \label{chirho}
\end{eqnarray}
Because of the averaging over the QCD partition function,
$\rho_c(\lambda_1,\lambda_2)$ depends on the quark masses.
Therefore it is not possible to derive the two-point correlation 
function by inverting the relation (\ref{chirho}). In order to invert 
this relation and to compute the correlation function one has to
introduce special scalar sources, unequivocally related to
the eigenvalues of the QCD Dirac operator, in the QCD partition
function. This partition function contains extra degrees of
freedom: one fermionic and one bosonic ghost-quark for each
special scalar source  related to an argument of the spectral  
correlation function.                   
The pqQCD partition function with two special sources
required for the calculation of the spectral two-point function is
given by
\begin{eqnarray}
Z^{\rm pqQCD}  &=& \int\! [dA]_{\nu} ~\frac{\det(i\Dirac +
z_1+j_1/2)}{\det(i\Dirac + z_1-j_1/2)} \frac{\det(i\Dirac +
z_2+j_2/2)}{\det(i\Dirac + z_2-j_2/2)}
\prod_{f=1}^{N_{f}} \det(i\Dirac + m_f) ~e^{-S_{\rm YM}[A]} \nonumber \\
&=& \left \langle \prod_k \Big( \prod_{f=1}^{N_f} (i \lambda_k+m_f)
\Big)
 \frac{i \lambda_k+z_1+j_1/2}{i \lambda_k+z_1-j_1/2} \;
\frac{i \lambda_k+z_2+j_2/2}{i \lambda_k+z_2-j_2/2} 
\right \rangle_{\rm YM}.\nonumber\\
\label{pqQCD}
\end{eqnarray}
Notice that for $j_1=j_2=0$ the pqQCD partition function reduces
to the QCD partition function. Therefore, the derivatives of the
pqQCD partition function with respect to the source terms at
$j_1=j_2=0$ are given by averages over the QCD partition function.
This enlarged partition function was first introduced to study the
quenched limit in QCD and is therefore known as partially quenched
QCD (pqQCD) \cite{pqChPT}. It has already been used to compute the
spectral density of the QCD Dirac operator in \cite{OTV,DOTV}. In
that case, only one such special scalar source had to be
introduced.

We will be interested in the disconnected scalar susceptibility
defined by
\begin{eqnarray}
  \label{schi}
  \chi (z_1,z_2) &=&\frac{1}{V} \; \partial_{j_1}
\partial_{j_2} \; {\rm ln}  Z_{\rm pqQCD}\Big|_{j_1=j_2=0} \\
  &=& \frac1V \left \langle \sum_k \frac 1{i \lambda_k+z_1} \; \sum_j
\frac 1{i \lambda_j+z_2}  \right\rangle_{\rm QCD} \nonumber \\ 
     &&-\frac1V \left \langle \sum_k \frac 1{ i \lambda_k+z_1}
\right \rangle_{\rm QCD} \;
 \left \langle \sum_j \frac 1{i \lambda_j+z_2} \right\rangle_{\rm QCD}.
\nonumber
\end{eqnarray}
It is related to the spectral two-point                       
correlation function of the QCD Dirac operator (\ref{rho2}) as follows 
\begin{eqnarray}
  \label{BC2}
  \chi(z_1,z_2)=\frac1V \int_{-\infty}^\infty
d\lambda_1 \; \int_{-\infty}^\infty d\lambda_2 \;
\frac{\rho_c(\lambda_1,\lambda_2)}{(i \lambda_1+z_1) (i
\lambda_2+z_2) } . \label{pqchirho}
\end{eqnarray}
Because the average over the QCD partition function does not
depend on $z_1$ and $z_2$, this integral equation can be inverted.
Using that $\chi(z_1,z_2)$ is odd in both $z_1$ and $z_2$, we find
that
\begin{eqnarray}
  \label{chidisc}
 \frac1V  \rho_c(\lambda_1,\lambda_2)&=&\frac 1{4 \pi^2}
{\rm Disc}\Big|_{z_1=i \lambda_1, \; z_2=i \lambda_2}
\chi(z_1,z_2)
\nonumber \\
&=&\frac1{4\pi^2} \lim_{\epsilon \rightarrow 0+} \Big( \chi(i
\lambda_1+\epsilon,i \lambda_2+\epsilon)+
\chi(-i \lambda_1+\epsilon,i \lambda_2+\epsilon)    \nonumber \\
&&+ \chi(i \lambda_1+\epsilon,-i \lambda_2+\epsilon)+ \chi(-i
\lambda_1+ \epsilon,-i \lambda_2+\epsilon) \Big).
\end{eqnarray}

At low energies, as already discussed in \cite{OTV,DOTV}, the
properties of the spectrum of the QCD Dirac operator can be
obtained from the low-energy limit of (\ref{pqQCD}). As is the
case for the usual chiral Lagrangian, this effective theory is
completely determined by the symmetries of the pqQCD partition
function. This topic will be discussed in the next section. An
alternative way to derive the perturbative results for the scalar
susceptibility is the use of the replica method
\cite{Berb99,Damsplitpq}. However, nonperturbative results cannot
be obtained this way
\cite{Vzirn,Kamenev,lerner,kanzieper,Z1,damsplitrep,Dalmazi,ADDV}.

Partially quenched Chiral 
Perturbation Theory, which is the basis of most of our results, 
will be introduced in section 2. The          
zero momentum sector of this theory will be analyzed in section 3.
It will be shown that in this domain the two-point spectral
correlation function for
 arbitrary topological  charge is  given by
chiral Random Matrix Theory. This nonperturbative result
generalizes a  calculation in \cite{Andreev} to arbitrary
topological charge without starting from chiral Random Matrix
Theory, but rather from an effective chiral Lagrangian which is
obtained from the symmetries of the microscopic 
theory. The contribution of 
the nonzero momentum modes to the scalar susceptibility and its
discontinuity is calculated in section 4. In section 5 we evaluate
the number variance for the different domains in the Dirac
spectrum. Concluding remarks are made in section 6.

\section{Partially Quenched Chiral Perturbation Theory}

It is well-known that the low-energy limit of QCD is given by a
theory of weakly interacting Goldstone bosons. The reason is
two-fold: the spontaneous breaking of chiral symmetry and the
existence of a mass gap for the non-Goldstone excitations. In QCD,
chiral symmetry is maximally broken consistent with the
Vafa-Witten theorem, implying that $SU_L(N_f) \times SU_R(N_f)$ is
broken spontaneously to the diagonal subgroup $SU_V(N_f)$. The
corresponding low-energy effective theory has been investigated in
great detail by means of Chiral Perturbation Theory. It describes
successfully the strong interaction phenomenology at low energies
\cite{Wein,GaL,Foundations}.

In this section we construct the low-energy limit of pqQCD.    
This low-energy effective theory, known as partially quenched
Chiral Perturbation Theory (pqChPT), is again solely based on the
symmetries of pqQCD and is given by a theory of Goldstone modes
associated with  the spontaneous symmetry breaking of chiral
symmetry. Because the unitary symmetry is not a good symmetry for
the bosonic ghost quarks, i.e., unitary transformations violate
the complex conjugation of the fields necessary to obtain a
convergent integral, we start from the complexified flavor
symmetry group given by $Gl(N_f+2|2) \times Gl(N_f+2|2)$. The
flavor symmetry group of the bosonic ghost quarks is then  given
by $Gl(2)/U(2) \times Gl(2)/U(2)$ which results in convergent
bosonic integrals. The natural choice for the flavor symmetry
group of the fermionic quarks  is $U(N_f+2)\times U(N_f+2)$. As is
the case for QCD we will assume that the axial symmetry of the
pqQCD is maximally broken by the vacuum expectation value of the
chiral condensate. We will also assume that supersymmetry is not
spontaneously broken. Because of spontaneous breaking of chiral
symmetry to the diagonal subgroup, the complete Goldstone
manifold, which also includes fermionic Goldstone modes, is then
given by the maximum super-Riemannian submanifold of $Gl(N_f+2|2)$
\cite{class,OTV,DOTV}. It will be denoted by             
$\widehat{Gl}(N_f+2|2)$. It 
consists of  a fermion-fermion block given by $U(N_f+2)$ and a
boson-boson block given  by $Gl(2)/U(2)$. The matrix elements of
the boson-fermion and the fermion-boson blocks of this manifold
are given by independent Grassmann variables. The unbroken chiral
symmetry group of the pqQCD partition function is thus given by
$\widehat{Gl}_L(N_f+2|2)\times \widehat{Gl}_R(N_f+2|2)$.

At low energies the relevant excitations are the Goldstone fields
parameterized by
\begin{eqnarray}
  U=\exp (i \sqrt{2}  \Pi^a T_a/F),
\end{eqnarray}
with $T_a$ the generators of the Goldstone manifold
$\widehat{Gl}(N_f+2|2)$. Under  a $\widehat{Gl}_L(N_f+2|2)
\times\widehat{Gl}_R(N_f+2|2)$ transformation of the quark fields,
the Goldstone fields transform as \be U \to U_L U U_R^{-1}. \ee
The low energy effective partition function is obtained from the
requirement that its  transformation properties are the same as
for the pqQCD  partition function. Under flavor transformations
the pqQCD partition function transforms as \be Z^{\rm pqQCD}(U_L
{\cal M} U_R^{-1}, \theta - i \log [{\rm Sdet} U_L U_R^{-1}]) =
Z^{\rm pqQCD}( {\cal M},\theta), \label{transzpq} \ee where
$\theta$ is the vacuum angle, and the quark mass matrix is denoted by 
${\cal M}$. To lowest order in the momenta and quark masses, the
effective partition function is thus given by \be Z^{\rm
eff}({\cal M}, \theta) = \int_{U \in \widehat{Gl}(N_f+2|2)} dU\,
 e^{-\int d^4 x {\cal L}(x)}
\ee with the effective chiral Lagrangian given by
\cite{pqChPT,OTV} \be {\cal L} =\frac{F^2}{4} {\rm
Str}(\partial_\mu U
\partial_\mu U^{-1})
- \frac{\Sigma}{2} {\rm Str}({\cal M}^\dagger U+ {\cal M} U^{-1})
+\frac{\Sigma \bar{m}}{2} (\frac{\Phi}{F}-\theta)^2.
\label{LpqChPT} \ee 
Here, the quark mass matrix is defined by            
\be
{\cal M}={\rm diag} (z_1+j_1/2,z_2+j_2/2, \underbrace{m, \cdots,
m}_{N_f}, z_1-j_1/2, z_2-j_2/2). 
\ee 
The axial anomaly is included
through a mass term for the super-$\eta'$  field,   $\Phi=-i F {\rm
Str} \ln U$, while integrating over the full axial symmetry group 
$\widehat{Gl}(N_f+2|2)$.
This term serves as a constraint that projects out the flavor
singlet channel. The first two terms in (\ref{LpqChPT}) also
appear in ChPT to lowest order. However, in the case of pqChPT,
there are both fermionic and bosonic Goldstone modes. Their masses
are given by $\sqrt{2 \Sigma m}/F$, $\sqrt{2 \Sigma z_1}/F$,
$\sqrt{2 \Sigma z_2}/F$, $\sqrt{\Sigma (m+z_1)}/F$, $\sqrt{\Sigma
(m+z_2)}/F$, and $\sqrt{\Sigma (z_1+z_2)}/F$, depending on their 
quark content.

The partially quenched effective partition function was first
formulated for the supergroup $U(N_f+k|k)$ to study the quenched
approximation in QCD \cite{pqChPT}. This formulation is suitable
for perturbative calculations. However, an effective partition
function based on this group cannot be used for nonperturbative
calculations of the group integral. For example, the supertrace
leads to the appearance of both positive and negative masses. The
correct integration manifold is the super-Riemannian manifold
$\widehat{Gl}(N_f+2|2)$ \cite{class,OTV,DOTV}.

As already mentioned in the introduction, it is possible to
distinguish different domains in the spectrum of the Dirac
operator. First, the effective partition function is only valid
for $z_1, z_2 \ll \Lambda_{\rm QCD}$. An important scale is given
by the Thouless energy defined as the quark mass scale for which
the Compton wavelength of the lightest corresponding Goldstone
mode is equal to the size of the box, i.e. \be \frac
{m_c\Sigma}{F^2} = \frac 1{L^2} . \ee For $|z_i| \ll m_c$ the
$z_i$-dependence of the condensate is determined by the
fluctuations of the zero momentum modes. In this domain, the
partition function factorizes into a zero momentum sector and a
nonzero momentum sector \cite{OTV,GL,LS}. In \cite{OTV,DOTV} it
was shown that in this domain the pqQCD partition function for the
one-point function reduces to the chRMT partition function. In
this article, we will show that the same is true for the spectral
two-point correlation function. Inside this domain, we can
distinguish a second scale: the smallest nonzero eigenvalue \be
\lambda_{\rm min} = \frac\pi{\Sigma V}. \ee For $|z_i| \gg
\lambda_{\rm min} $ the group integrals can be evaluated
perturbatively, whereas for smaller values of the $z_i$ the group
integrals have to be calculated exactly. In the domain $z_i \geq
F^2/\Sigma L^2$, the nonzero momentum modes become important and
have to be taken into account. These three domains for the
spectral two-point function will be analyzed in detail in the
remainder of this article.

\section{Quantum and Ergodic domains}
In these domains, corresponding to $|z_i| \ll m_c$, the zero
momentum mode sector and the nonzero momentum sectors of the
pqChPT partition function factorize \cite{OTV}. The
$z_i$-dependence of the partition function and therefore the
spectral two-point function can be obtained from its zero momentum
part only. It is given by the super-unitary matrix integral \be
Z_{\rm eff}({\cal M}, \theta) = \int_{U   \in
\widehat{Gl}(N_f+2|2)} dU
 e^{
\frac{\Sigma V}{2} {\rm Str}({\cal M}^\dagger U+ {\cal M} U^{-1})
-\frac{\Sigma \bar{m}V}{2} (\frac{\Phi}{F}-\theta)^2}. \ee We
decompose the partition function according to \be Z_{\rm
eff}({\cal M}, \theta) = \int d\nu e^{i\nu\theta} Z_\nu({\cal M}),
\ee 
with the partition function in the sector of topological
charge $\nu $ given by
(for an in depth analysis of the normalization factors of the 
different topological sectors we refer to \cite{poultop})      
\be Z_\nu ({\cal M}) &=&
 C_1 e^{-\frac{\nu^2}{2 \bar m \Sigma V}}
\int_{U \in \widehat{Gl}(N_f+2|2)} dU\,
 {\rm Sdet}^\nu U
e^{\frac{\Sigma V}{2} {\rm Str}({\cal M} U+ {\cal M}^\dagger
U^{-1})}. \label{ZpqChPT} \ee Exact integrals over the Goldstone
manifold $\widehat{Gl}(N_f+2|2)$ are mathematically quite
complicated. In the following we will restrict ourselves to the
simplest case, and calculate the scalar susceptibility in the
quantum and ergodic domains only in the quenched limit. A
calculation of the two-point correlation function based on the
supersymmetric formulation of chiral Random Matrix Theory
\cite{Efetov,VWZ} was first given in \cite{Andreev} for the sector
of zero topological charge. In this article we start directly from
the partially quenched chiral Lagrangian and extend the
calculation to all values of the topological charge. 
Remarkably, the $k$-point correlation functions can be expressed in 
terms of the zero momentum part of the ChPT partition function with 
$2k$    
additional flavors \cite{FV}. The relation of these additional     
flavors with the equal number of                                   
additional flavors that enter in the pqQCD partition has been      
worked out in detail for the one-point function \cite{OTV}.        

\subsection{Scalar Susceptibility in the Quenched Limit}

In the quenched limit and in a sector of topological charge $\nu$,
the zero-mode part of the effective partition function can be
written as:
\begin{eqnarray}
  \label{PartFct}
  Z^\nu_{\rm eff}(z_1,z_2,j_1,j_2)=\int_{\widehat{Gl}(2|2)} DU \; 
{\rm Sdet}^\nu                                               
  U \; \exp \{
  \frac{\Sigma V}2  {\rm Str} {\cal M} (U+U^{-1}) \},
\end{eqnarray}
where the mass matrix is given by
\begin{eqnarray}
  \label{}
  \Sigma V {\cal M}=\left( \begin{array}{cc} M_1 & 0 \\ 0 & M_2 \end{array}
\right),
\end{eqnarray}
with $M_i=\Sigma V {\rm diag}(z_i+j_i/2,z_i-j_i/2)$. The Goldstone
manifold $\widehat{Gl}(2|2)$ is the maximum Riemannian submanifold
of ${Gl}(2|2)$ with fermion-fermion block given by $U(2)$ and
boson-boson block given by $Gl(2)/U(2)$ \cite{class,OTV,DOTV}.    

\subsubsection{Parameterization of the Goldstone manifold}

We parameterize the Goldstone manifold in terms of Goldstone modes
related with the one-point functions corresponding to $z_1$ and
$z_2$, and Goldstone modes that describe  two-point correlations.
A convenient parameterization is given by \cite{Andreev},
\begin{eqnarray}
  \label{Uparam}
  U= \left( \begin{array}{cc}  w_1 & 0  \\ 0  & w_2  \end{array} \right)
     \left( \begin{array}{cc}  \sqrt{1-w \bar w} & w \\ - \bar w  &
     \sqrt{1-\bar w w} \end{array} \right)
     \left( \begin{array}{cc}  w_1 & 0  \\ 0  & w_2 \end{array} \right),
\end{eqnarray}
where  $ w_{1,2} \in \widehat{Gl}(1|1)$ and $w,\,  \bar w \in
Gl(1|1)$. An explicit parameterization of these supermatrices will
be given below.
The advantage of the parameterization (\ref{Uparam}) becomes clear
upon consideration of the supertrace that appears in the exponent
in the effective partition function (\ref{PartFct}):
\begin{eqnarray}
  \label{STR}
  \frac{\Sigma V}2 {\rm Str} {\cal M} (U+U^{-1})={\cal S}_1+{\cal S}_2,
\end{eqnarray}
with
\begin{eqnarray}
  \label{Str1}
   {\cal S}_1=\frac12 {\rm Str} \Big[ M_1 \Big(w_1
\sqrt{1-w \bar w} w_1+w_1^{-1} \sqrt{1-w \bar w} w_1^{-1} \Big)
\Big],
\end{eqnarray}
and
\begin{eqnarray}
  \label{Str2}
  {\cal S}_2=\frac12 {\rm Str} \Big[M_2 \Big(w_2
\sqrt{1-\bar w w} w_2+w_2^{-1} \sqrt{1- \bar w w} w_2^{-1} \Big)
\Big].
\end{eqnarray}
Furthermore, the source terms $j_1$ and $j_2$ only occur in
combination with $w_1$ and $w_2$, respectively. As we will see in
the next subsection, the invariant measure of $\widehat{Gl}(2|2)$
in the parameterization (\ref{Uparam}) factorizes according to \be
\mu(w,\bar w ,w_1, w_2) = \mu(w, \bar w ) \mu(w_1) \mu(w_2). \ee
We thus find that the integrals over $w_1$ and  $w_2$ in the
computation of the scalar susceptibility factorize
\begin{eqnarray}
  \label{susc}
  \chi(z_1,z_2)=\frac1V \frac{\partial^2}{\partial j_1 \partial j_2} \log Z
\Big|_{j_1=j_2=0}=\frac1V \int dw d \bar w \; \mu(w, \bar w) \;
{\cal I}_1(w,\bar w) {\cal I}_2(w,\bar w),
\end{eqnarray}
where the  integrals ${\cal I}_1(w,\bar w) $ and ${\cal
I}_2(w,\bar w ) $ are given by
\begin{eqnarray}
  \label{1pt}
  {\cal I}_i=\int dw_i  \; \mu(w_i) \; \partial_{j_i}
 {\cal S}_i \; e^{ {\cal S}_i} \Big|_{j_i=0}, \; \; i=1,2.
\end{eqnarray}

A further simplification arises by using a polar decomposition of
the $2\times 2$ supermatrices that appear in the parameterization
(\ref{Uparam}), \be
  \label{Gdiag}
         w_i&=&v_i \Lambda_i v_i^{-1}, \quad i=1,2,  \\
         w&=&v S u^{-1}, \qquad   \bar w = u \bar S v^{-1},
\ee where $\Lambda_{1,2}$ and $S$ are $2 \times 2$ diagonal
supermatrices with commuting elements given by \be
  \label{diagonal}
  \Lambda_i&=&{\rm diag} (e^{i \psi_i/2}, e^{s_i/2} ), \quad i=1,2, \\
     S&=& {\rm diag} (\sin \theta e^{i \rho}, i\sinh \phi  e^{i \sigma} ), \\
\bar S&=& {\rm diag}
    (\sin \theta e^{-i \rho}, i\sinh \phi  e^{-i \sigma} ),\\
  C&=& \sqrt{1-S \bar S}={\rm diag} (\cos \theta , \cosh \phi),
\ee and $u$, $v$, $v_1$, and $v_2$, all elements of
$U(1|1)/U(1)\times U(1)$, can be conveniently parameterized
according to \be u &=&\exp \mat 0 & \zeta \\ \chi &0 \emat, \qquad
v =\exp \mat 0 & \xi   \\ \eta & 0 \emat,  \nonumber \\
v_i& =&\exp \mat 0 & \xi_i \\ \eta_i & 0 \emat, \quad i=1,2.
  \label{angles}
\ee After these ``Grassmann diagonalizations'', the supertraces in
(\ref{Str1}) and  (\ref{Str2}) are given by 
\begin{eqnarray}
  \label{Str1bis}
  {\cal S}_1=\frac12 {\rm Str} \Big[ v_1^{-1} v C v^{-1} v_1
\Big( \Lambda_1 v_1^{-1} M_1 v_1 \Lambda_1+ \Lambda_1^{-1}
v_1^{-1} M_1 v_1 \Lambda_1^{-1} \Big) \Big],
\end{eqnarray}
and
\begin{eqnarray}
  \label{Str2bis}
  {\cal S}_2=\frac12 {\rm Str} \Big[ v_2^{-1} u C u^{-1} v_2
\Big( \Lambda_2 v_2^{-1} M_2 v_2 \Lambda_2+ \Lambda_2^{-1}
v_2^{-1} M_2 v_2 \Lambda_2^{-1} \Big) \Big].
\end{eqnarray}

\subsubsection{Measure}

The parameterization of the Goldstone manifold is of the form \be
U = W T W = W^2 W^{-1} T W \equiv W^2 T'. \ee The Berezinian of
the transformation from the variables $T'$ to $T$ is one. To
calculate the invariant measure we thus consider \be T' [U^{-1}
dU] {T'}^{-1} = W^{-2}dW^2 + [dT']\, {T'}^{-1}. \ee The measure
thus factorizes into a product of one factor depending only  on
$W$ and another factor depending only on $T'$ (and on $T$ after
the transformation from $T'$ to $T$). The $W$-dependent part of
the measure trivially factorizes into a $w_1$-dependent piece and
a $w_2$-dependent piece. We thus find \be
d\mu(U) &=& w_1^{-2} d w_1^2\, w_2^{-2} d w_2^2\, T^{-1} dT \nonumber \\
        &\equiv& \mu(w_1) dw_1 \, \mu(w_2) dw_2 \, \mu(w,\bar w) dw d\bar w.
\ee The first two integration measures simply follow from the
invariant measure of $Gl(1|1)/U(1|1)$ whereas the integration
measure of the $T$-integrations is given by  the invariant measure
of $Gl(1|1)$. Both measures will be calculated in the next part of
this section.

The matrices $w_i$ in the coset $Gl(1|1)/U(1|1)$ have four independent 
parameters and can be parameterized as in (\ref{Uparam},
\ref{angles}). We first calculate the measure $w_i^{-1} d w_i$. To
obtain the measure $w_i^{-2} d w_i^2$ we only have to replace the
diagonal elements by their square. The invariant measure is given
by the Berezinian of the transformation from variables \be \delta
w'_i \equiv v^{-1}_i[ w^{-1}_i dw_i ]v_i \ee to variables
$d\psi_i$, $ds_i$, $d\xi_i$ and $d\eta_i$. One easily derives that
\be \delta w'_i = \Lambda^{-1}_i\delta w''_i = \Lambda^{-1}_i[
\delta v_i \Lambda_i +  d \Lambda_i - \Lambda_i \delta v_i], \ee
where $\delta v_i =v^{-1}_i dv_i$. The Berezinian of the 
transformation from the variables $\delta w''_i$ to the $d\Lambda_i$ 
and the offdiagonal elements of $\delta v_i$ is given by \be
 &&{\rm Sdet}  \matf 1  &  0  &  0  &  0 \\
                    0  &  1  &  0  &  0 \\
                    0  &  0  & e^{s_i/2}-e^{i\psi_i/2}  &  0 \\
                    0  &  0  &    0  & e^{i\psi_i/2}-e^{s_i/2} \ematf
                \nonumber \\
  &=& \frac 1{(e^{s_i/2}-e^{i\psi_i/2})(e^{i\psi_i/2}-e^{s_i/2})}.
\ee The Berezinian of the transformation from $\delta w''$ to
$\delta w'$ is one (factors from the bosonic integrations cancel
against factors from the fermionic integrations), and $(\delta
v)_{12}(\delta v)_{21} = d\xi d\eta$. For the Berezinian, denoted
by $B$, we thus find \be w_i^{-1} dw_i \equiv Bd\psi_i ds_i d\xi_i
d \eta_i &=& \frac{ de^{s_i/2} de^{i\psi_i/2} d\xi_i
d\eta_i}{(e^{s_i/2}-e^{i\psi_i/2})(e^{i\psi_i/2}-e^{s_i/2})} \nonumber \\
&=& \frac{i ds_i d\psi_i  d\xi_i
d\eta_i}{4(e^{s_i/2}-e^{i\psi_i/2})(e^{-s_i/2}-e^{-i\psi_i/2})}.
\ee The integration measure for the $w_i$ variables is then simply
obtained by squaring the diagonal elements of $\Lambda_i$, \be
\mu(w_i) dw_i = \frac{i ds_i d\psi_i  d\xi_i
d\eta_i}{(e^{s_i}-e^{i\psi_i})(e^{-s_i}-e^{-i\psi_i})}.
\label{measurewi} \ee

Next we calculate the invariant measure of $Gl(1|1)$. This group
has eight independent parameters and can be parameterized as in
(\ref{Uparam}) and (\ref{angles}). With the observation that 
\cite{Vzirn,lerner}  $T^{-1}dT = dw d \bar w$ we calculate the 
integration measure starting from \be
dw' &\equiv& v^{-1} dw u = \delta v S + dS - S \delta u, \nonumber \\
d \bar w' &\equiv& u^{-1} d \bar w u = \delta u \bar S + d\bar S -
\bar S \delta v, \ee with $\delta u = u^{-1} du$ and $\delta v =
v^{-1} dv$. The Berezinian of the transformation from the
variables on the left hand side to the variables on the right hand
side is given by \be \frac 1{(\sin^2\theta +\sinh^2 \phi)^2}
=\frac 1{(\cos^2\theta -\cosh^2 \phi)^2}. \ee The Jacobian for the
transformation of  the variables $\{S_{11},\, S_{22},\,
S_{11}^*,\, S_{22}^*\}$ to the variables $\{\theta,\, \phi,\,
\rho,\, \sigma\}$ is simply given by \be i\sinh 2\phi \sin
2\theta. \ee For the invariant measure of $Gl(1|1)$ we thus find 
\be T^{-1}
dT = \mu(w,\bar w )dw d \bar w = \frac{i\sinh 2\phi \sin 2\theta
}{(\cos^2\theta -\cosh^2 \phi)^2} d\theta d\phi d\rho d\sigma
d\zeta d \chi d \xi d \eta. \label{measurew} 
\ee

\subsubsection{Integration over the Goldstone manifold}

The first supertrace ${\cal S}_1$ in (\ref{Str1bis}) appearing in
the integral (\ref{susc}) can be written as
\begin{eqnarray}
 \frac{{\cal S}_1}{\Sigma V}&=&(z_1+\frac12 j_1) \cos \theta \cos \psi_1
- (z_1-\frac12 j_1)
   \cosh \phi \cosh s_1 \nonumber \\
   &+& j_1 (\cos \theta \cos \psi_1-\cosh \phi \cosh s_1) \xi_1 \eta_1  \\
   & +&(\cos \theta-\cosh \phi) \Big( (z_1+\frac12 j_1) \cos \psi_1 -
   (z_1-\frac12 j_1) \cosh s_1 \Big) (\xi-\xi_1) (\eta-\eta_1) \nonumber \\
   & +&j_1 \cos \psi_1 \cosh s_1 (\cos \theta-\cosh \phi)
   \Big( (\xi-\xi_1) \eta_1+ \xi_1 (\eta-\eta_1)   \Big) \nonumber \\
   & +&j_1 (\cos \theta-\cosh \phi) (\cos \psi_1-\cosh s_1) \xi \eta
   \xi_1 \eta_1. \nonumber
\label{Sexplicit}
\end{eqnarray}
The Grassmann integrals in (\ref{susc}) are calculated by
collecting the coefficient of $\xi \eta \xi_1 \eta_1$ in the
expansion of $e^{{\cal S}_1}$. Only the terms linear in $j_1$
contribute to susceptibility. Using the nilpotency of the
Grassmann variables, such as for example $(\xi -\xi_1)^2 =0$, one
easily shows that 
\be 
\frac1{\Sigma V} \int d\xi d\eta d\xi_1 d\eta_1 \partial_{j_1} e^{{\cal 
S}_1}|_{j_1=0} &= & e^{\Sigma V z_1                                     
(\cos\theta \cos \psi- \cosh \phi \cosh s)}                             
(\cos\theta -\cosh \phi)(\cos \psi_1 - \cosh s_1)\nonumber \\
&\times& (1 + \Sigma V z_1 \cos\theta \cosh \psi_1-\cosh\phi \cosh
s_1). 
\ee 
An analogous result is obtained from the  expansion of
$e^{{\cal S}_2}$. The integration over the Grassmann variables in
the disconnected scalar susceptibility (\ref{susc}) is now
trivial. Taking into account the measures  (\ref{measurew},
\ref{measurewi}) we arrive at
\begin{eqnarray}
  \label{suscGrass}
  \chi(z_1,z_2)=4{\Sigma^2 V} \int_0^1 dt \int_1^\infty dp \;
\frac{tp}{(t^2-p^2)^2} \; {\cal F}(\Sigma V z_1) {\cal F}(\Sigma V
z_2),
\end{eqnarray}
where
\begin{eqnarray}
  {\cal F}(x)
 &=& (t-p) \int_0^{2 \pi} d \psi \int_0^\infty d s \;
\frac1{(e^{i \psi}-e^{s})(e^{-i \psi}-e^{-s})}  \; e^{x (t \cos
\psi-p \cosh s)} \;
e^{\nu (i \psi-s)}  \nonumber \\
&\times &(\cos \psi-\cosh s) \Big(1+ x (t \cos \psi-p \cosh s)
\Big).
\end{eqnarray}
To avoid Efetov-Wegner terms \cite{GPW,Zirnhal} and problems
related to the singularity in the integrand of ${\cal F}(x)$, we
compute
\begin{eqnarray}
  \label{Deriv1pt}
  \frac 1{ (t-p)} \Big( \frac \partial {\partial t}+\frac \partial
{\partial p}  \Big) {\cal F}(x)&=&- \frac{x}4 \int_0^{2 \pi} d
\psi \int_0^\infty d s \;
e^{x(t \cos \psi-p \cosh s)} \; e^{\nu (i \psi-s)}  \nonumber \\
&\times&(1-e^{s +i\psi})(1-e^{s -i\psi})
\Big( 2+x  (t \cos \psi-p \cosh s) \Big) \nonumber \\
&=&  \Big( \frac \partial {\partial t}+\frac \partial {\partial p}
\Big) \;  \Big( (t+p) I_\nu(x t) K_\nu (x p)  \Big).
\end{eqnarray}
In the first equality we have used the identity \be \frac{\cos
\psi - \cosh s}{e^{i\psi} - e^s} = \frac 12 ( 1 - e^{-s- i\psi}),
\ee and the second equality follows from relations for modified
Bessel functions. We thus  find that
\begin{eqnarray}
  {\cal F}(x)=(t^2-p^2) I_\nu(x t) K_\nu (x p).
\end{eqnarray}
Inserting this expression into (\ref{suscGrass}), 
the final result for the                                  
disconnected scalar susceptibility is given by
\begin{eqnarray}
  \chi(z_1,z_2)&=& 4\Sigma^4 V^3 z_1 z_2 \int_0^1 dt\, t
I_\nu(\Sigma V z_1 t) I_\nu(\Sigma V z_2 t) \int_1^\infty dp\, p
K_\nu(\Sigma V z_1 p)
K_\nu(\Sigma V z_2 p) \nonumber \\
&=&\frac{4 \Sigma^2 V z_1 z_2}{ (z_1^2-z_2^2)^2} \Big( z_1
I_{\nu+1}(\Sigma V z_1) I_\nu(\Sigma V z_2)- z_2 I_{\nu+1}(\Sigma
V z_2) I_\nu(\Sigma V z_1) \Big)
  \label{suscFinal}  \\
&\times&\Big( z_1  K_{\nu+1}(\Sigma V z_1) K_\nu(\Sigma V z_2)-
z_2 K_{\nu+1}(\Sigma V z_2) K_\nu(\Sigma V z_1) \Big). \nonumber
\end{eqnarray}
In the limit $z_1=z_2$, this expression coincides with the result
obtained in \cite{Gock}.

\subsection{Two-point correlation function}
Finally, the two-point spectral correlation function is given by
the discontinuity of the disconnected scalar susceptibility across
the imaginary axis (\ref{chidisc}). This follows by using
relations for Bessel functions such as \be
K_\nu(iz) &=& \frac{\pi i}2 e^{\pi i \nu/2}(J_\nu(-z) +iN_\nu(-z)),\nonumber\\
K_\nu(-iz) &=& e^{-\pi i\nu} K_\nu(iz) - \pi i I_\nu(iz),\nonumber \\
I_\nu(iz) &=& e^{-\pi i\nu/2} J_\nu(-z), \qquad -\pi <{\rm arg}\,
z \le
\frac \pi 2,\nonumber \\
 -\frac 2{\pi z}&=&J_\nu(z)N_{\nu+1}(z) - J_{\nu+1}(z)N_\nu(z).
\ee In the quantum and ergodic domains, we then obtain the
two-point correlation function
\begin{eqnarray}
  \label{CorrelQuantum}
  \frac{\rho_c(\lambda_1,\lambda_2)}{V^2 \Sigma^2}&=&
\delta(\lambda_1  -\lambda_2 ) \frac {\lambda_1}2 [J^2_\nu(\Sigma
V \lambda_1)
- J_{\nu+1}(\Sigma V \lambda_1)J_{\nu-1}(\Sigma V \lambda_1 )] \\
&-&
 \frac{\lambda_1 \lambda_2}
{(\lambda_1^2-\lambda_2^2)^2} \Big( \lambda_1 J_{\nu+1}(\Sigma V
\lambda_1) J_\nu(\Sigma V \lambda_2) - \lambda_2 J_{\nu+1}(\Sigma
V \lambda_2) J_\nu(\Sigma V \lambda_1) \Big)^2.\nonumber
\end{eqnarray}
The first term represents the contribution due to the
selfcorrelations of the eigenvalues (see eq. (\ref{rho2})).  
It comes from the non-trivial
$\epsilon \rightarrow 0$ limit in the discontinuity of the
susceptibility (\ref{chidisc}). This results coincides with the
chRMT result \cite{VZ}. In the ergodic domain, where $\Sigma V    
\lambda_{1,2} \gg 1$, the two-point spectral correlation function 
reduces to
\begin{eqnarray}
  \label{CorrelErgodicRMT}
  \rho_c(\lambda_1,\lambda_2)=-\frac{1}{2\pi^2} \frac{1} 
{(\lambda_1-\lambda_2)^2}+\cdots,
\end{eqnarray}
and agrees with the asymptotic result for the two-point
correlation function of the Wigner-Dyson ensembles.

\section{Ergodic and Diffusive domains}

In these domains, corresponding to $\lambda_{\rm min} \ll |z_i|
\ll \Lambda_{\rm QCD}$, the scalar susceptibility can be computed
perturbatively within pqChPT with Lagrangian for $\theta =0$ given
by (\ref{LpqChPT}),
\begin{eqnarray}
  \label{Leff}
  {\cal L}_{\rm eff}=\frac{F^2}{4} {\rm Str}(\partial_\mu U
\partial_\mu U^{-1})
- \frac{\Sigma}{2} {\rm Str}({\cal M} (U+ U^{-1})) +\frac{M_0^2}2
\Phi^2
+\frac{\alpha}{2} \partial_\mu \Phi \partial_\mu \Phi. \nonumber \\
\end{eqnarray}
The $(N_f+4) \times (N_f+4)$ matrix $U$ is  parameterized as
\begin{eqnarray}
  U=\exp (i \sqrt{2}  \Pi^a T_a/F),
\end{eqnarray}
with $T_a$ the generators of the Goldstone manifold
$\widehat{Gl}(N_f+2|2)$, and the singlet field is normalized as
\begin{eqnarray}
\Phi =-i F
{\rm Str} \ln U.
\end{eqnarray}
Its mass is related to the topological susceptibility by
\cite{OTV} \be M_0^2 = \frac {2 \langle \nu^2 \rangle } {F^2 V}
\equiv \frac{ 2 \bar m \Sigma}{F^2}, \label{topsusc} \ee where
$\bar m$ has been introduced to simplify expressions below. For
$\bar m =m/N_f$ this expression gives the topological
susceptibility for $N_f$ light quarks with mass $m$, \be \langle
\nu^2 \rangle = \frac {m V \Sigma}{N_f}. \ee

We can distinguish two types of Goldstone modes: those that
correspond to the diagonal generators and those that correspond to
the off-diagonal generators. To one-loop order, the off-diagonal
Goldstone modes do not mix with the super-singlet field $\Phi$.
Therefore, their propagator is simply given by
\begin{eqnarray}
  G(p^2)=(p^2+M^2)^{-1},
\label{offdiagProp}
\end{eqnarray}
where $M$ is the mass of the corresponding Goldstone mode. The
diagonal Goldstone modes, on the other hand, do mix with the
super-singlet mode. It is still possible to diagonalize the
quadratic form  in the Goldstone fields (see
\cite{Golleung,pqChPT,TV}). This results in the following
propagator for diagonal mesons in the sector of fermionic quarks
($1 < i,j < N_f+2$),
\begin{eqnarray}
  G_{ij}(p^2)=\delta_{ij} \frac1{p^2+M^2_{ii}} - \frac{(\alpha p^2 + M_0^2)
(p^2+M^2)}{(p^2+M_{ii}^2) (p^2+M_{jj}^2)
((1+N_f \alpha) p^2+ M^2+N_f M_0^2)}, \nonumber \\
\end{eqnarray}
where $i$ denotes the flavor of the quarks of one of the diagonal
mesons, and $j$ the flavor of the quarks of the other one. The
masses in the propagator are given by the Gell-Mann--Oakes--Renner
relation, $M^2_{ii}=2 \Sigma z_i/F^2$ for $i=1,2$, $M^2\equiv
M^2_{ii}=2 \Sigma m/F^2$ for $i=3, \cdots N_f+2$.


We will consider two limits of this theory:  $M\to \infty$ (i.e.
$m\to \infty$), which is the quenched case, and $M_0 \to \infty$,
which is QCD with $N_f$ light flavors of quarks (and a completely
decoupled singlet field). In the quenched case, the propagator
matrix for the ``diagonal mesons'' is given by
\begin{eqnarray}
  \label{sprop}
   G_{ij}(p^2)=\delta_{ij} \frac1{p^2+M^2_{ii}} - \frac{(\alpha p^2 + M_0^2)
}{(p^2+M_{ii}^2) (p^2+M_{jj}^2) }, \quad {\rm for}\quad 1 \le i,j
\le 2,
\end{eqnarray}
whereas in the QCD limit we find
\begin{eqnarray}
\label{sprop2}
 G_{ij}(p^2)=
\delta_{i j} \frac1{p^2+M^2_{ii}} - \frac{(p^2+M^2)}{N_f
(p^2+M_{ii}^2)
 (p^2+M_{jj}^2)}, \quad {\rm for}\quad 1\le i,j \le N_f +2.
\end{eqnarray}
Below we only need the propagators for $1\le i,j \le 2$ so that
both limits can be treated at the same time by introducing the
propagator \be
   G_{ij}(p^2)=\delta_{ij} \frac1{p^2+M^2_{ii}} - \frac{(\eta p^2 + 2\Sigma
m_\eta/F^2) }{(p^2+M_{ii}^2) (p^2+M_{jj}^2) }, \quad {\rm
for}\quad 1 \le i,j \le 2. \label{diagProp} \ee In the QCD limit
we have that $\eta = 1/N_f$ and $m_\eta = m/N_f$, whereas the
quenched case is obtained by the substitution $\eta =\alpha$ and $
m_\eta  =\bar m$.

\subsection{Disconnected scalar susceptibility}

We compute the disconnected scalar susceptibility for spectral
quarks with masses given by $z_1$ and $z_2$. The mass of all $N_f$
other quarks is taken to be equal to $m$. To one-loop order the
contributions represented by the following diagrams have to be
taken into account:
\newline
\begin{center}
\begin{figure}[!ht]
\hspace*{4cm}
\epsfig{file=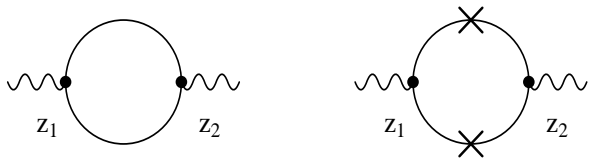,height=2.2cm,width=8cm,clip=}
\caption[]{\small One-loop diagrams in pqChPT which contribute to
$\chi(z_1,z_2)$. The full lines represent either the ``standard'' 
propagator of  an off-diagonal meson (\ref{offdiagProp}) or the
propagator of a diagonal meson (\ref{diagProp}) (with a cross).
The wiggly lines denote the two different scalar sources.}
\end{figure}
\end{center}

\noindent The propagators in the first diagram in Fig. 1 are given
by $G(p^2)=(p^2+M^2_{12})^{-1}$  with $M_{12}^2= \Sigma
(z_1+z_2)/F^2$. In the second diagram of Fig. 1, the lines with a
cross denote the propagators that mix the diagonal mesons with the
super-$\eta'$, $G_{12}(p)$ in (\ref{diagProp}). Such type of
contributions  also essential in the calculation of the
resolvent  from the partially quenched chiral Lagrangian as well  
(\cite{pqChPT,OTV,TV}). The  disconnected scalar susceptibility
computed to one-loop order within pqChPT is thus given by
\begin{eqnarray}
  \label{pertsusc}
  \chi(z_1,z_2)&=&\frac{\Sigma^2}{VF^4}  \sum_p  \Big[G(p^2)^2+
 2 G_{12}(p^2)^2 \Big] \nonumber \\
 &=&\frac{\Sigma^2}{V F^4}  \sum_p \Big[ \frac1{(p^2+M^2_{12})^2} +
 2\frac{(\eta p^2+2\Sigma  m_\eta /F^2)^2}
{(p^2+M^2_{11})^2 (p^2+M^2_{22})^2} \Big].
\end{eqnarray}
In the notation of \cite{HasLeu} with \be \label{GI}
G_r(M^2)=\frac{\Gamma(r)}{V} \sum_p (p^2+M^2)^{-r}, \ee the scalar
susceptibility can be rewritten as
\begin{eqnarray}
\label{suscfirst}
\chi(z_1,z_2)&=&\frac{\Sigma^2}{F^4} \left[ G_2(M^2_{12})+\frac{2
(\eta z_1-m_\eta)^2}{ (z_1-z_2)^2} G_2 (M^2_{11}) +\frac{2 (\eta
z_2-m_\eta)^2}{(z_1-z_2)^2}
G_2 (M^2_{22})  \right.   \\
&&\left. \hspace{2cm}+\frac{F^2}{\Sigma} \frac{2 (\eta z_1-m_\eta)
(\eta z_2-m_\eta)}{ (z_1-z_2)^3} \Big( G_1(M_{11}^2)-
G_1(M_{22}^2) \Big) \right] .\nonumber
\end{eqnarray}

\noindent The functions $G_1(M^2)$ and $G_2(M^2)$ for momenta in a
finite box were analyzed in detail in \cite{HasLeu}. They are
obviously related to properties of the propagator of a free scalar
particle at the origin (notice that $G_2(M^2)=-\partial_{M^2}
G_1(M^2)$). For a box with volume $L^4$ and  momentum cutoff
$\Lambda$ they are given by \cite{GaL,HasLeu}, \be
G_1(M^2)&=&\frac{1}{16 \pi^2} M^2 \log \frac{M^2}{\Lambda^2}+
g_1(M^2,L),
\nonumber\\
G_2(M^2)&=&-\frac{1}{16 \pi^2}(1+ \log\frac{M^2}{\Lambda^2})
+g_2(M^2,L), \label{deltag} \ee where \be g_r(M^2,L) = \frac
1{16\pi^2} \int_0^\infty d\lambda \lambda^{r-3} \sum_{{\bf n}\ne
0} e^{-\lambda M^2 - {\bf n}^2 L^2/4\lambda}, \ee and the sum is
over a four dimensional lattice of integers. The functions
$g_r(M^2)$ obviously vanish in the thermodynamic limit. For $
1/\Lambda L^2 \ll M \ll  1/L$ they dominate the logarithmic terms
in the propagators and  can be expanded in powers of $M$ resulting
in \be
g_1(M^2,L) &=& \frac{1}{M^2 L^4}+O(M^0/L^2), \nonumber\\
g_2(M^2,L) &=&\frac{1}{M^4 L^4}+ O(\log(ML)). \label{gexpansion}
\ee In the domain  $ 1/\Lambda L^2 \ll M_{ii} \ll  1/L$ ($i=1,2$)
the disconnected scalar susceptibility (\ref{pertsusc}) is thus
given by \be \chi(z_1,z_2)=\frac{m^2_\eta}{2 \; z_1^2 \; z_2^2 \;
V} +\frac{1}{(z_1+z_2)^2 V}+ \dots. \label{sus01} \ee Therefore in
the quenched limit, using the relation (\ref{topsusc}), the result
reads,
\begin{eqnarray}
  \chi(z_1,z_2)=\frac{\langle \nu^2 \rangle^2}{2 \; z_1^2 \; z_2^2 \;
\Sigma^2 \; V^3} +\frac{1}{(z_1+z_2)^2 V}+ \dots, \label{sus02}
\end{eqnarray}
and in the QCD limit, with $m_\eta = m/N_f$, one finds \be
\chi(z_1,z_2)=\frac{m^2}{2 N_f^2 \; z_1^2 \; z_2^2 \; V}
+\frac{1}{(z_1+z_2)^2 V}+ \dots. \label{sus03} \ee In the QCD
limit, for spectral quarks with equal masses $z_1=z_2=m$, we
indeed recover the ChPT result that can be easily derived from the
results given in \cite{SmilVerba}, \be \chi=\frac{N_f^2+2}{4
N_f^2} \frac{1}{m^2 V}+\dots. \ee

In the thermodynamic limit at fixed values of the quark masses,
 the finite volume corrections
$g_{1,2}(M^2,L)$ in (\ref{deltag}) can be ignored. This results in
the susceptibility
\begin{eqnarray}
 \label{thermsusc}
\chi(z_1,z_2)&=&-\frac{\Sigma^2}{16 \pi^2 F^2} \Bigg[ \ln
\frac{z_1+z_2}{2 \mu} \nonumber \\
&+&\frac{2}{(z_1-z_2)^3} \Bigg\{ \Big( (\eta z_1-m_\eta) [m_\eta
(z_1+z_2) +\eta z_1(z_1- 3z_2)] \Big) \ln \frac{ z_1}{\mu} -  z_1
\leftrightarrow     z_2  \nonumber \\
&+&(z_1-z_2) \Big( \eta^2 (z_1^2+z_2^2)-2 \eta m_\eta (z_1+z_2) +2
m_\eta^2  \Big)  \Bigg\} \Bigg],
\end{eqnarray}
where we have defined the
scale $\mu=\Lambda^2 F^2/2 \Sigma$ (compare to (\ref{deltag})). If
the two spectral quarks have the same mass $z_1=z_2=z$, a somewhat
simpler expression is obtained, \be \label{eqz}
\chi(z)=\frac{\Sigma^2}{16 \pi^2 F^4} \Bigg[ \frac{m_\eta^2}{3
z^2} +\frac{4 \eta m_\eta}{3 z} - \Big(1+ 2 \eta^2 \Big) \ln
\frac{z}{\mu}  \Bigg]. \ee In the limit $z \rightarrow 0$, the
disconnected scalar susceptibility is singular (notice that this
expression is only valid for $z \gg 1/V\Sigma$). This is due to
the double pole occurring in the neutral meson propagator when
$m_\eta \neq 0$. Such singularities are common in pqChPT. They
also appear, for instance, in the small mass behavior of the
scalar  radius of the pion in the quenched approximation
\cite{ColPal}.
In the case $m_\eta=0$, i.e. for $\langle \nu^2 \rangle=0$ or in
the chiral limit, the algebraic singularities in (\ref{thermsusc})
cancel and only a logarithmic singularity remains.

The physically more interesting QCD limit of the disconnected
scalar susceptibility, $\chi(z_1,z_2)$, is obtained from
(\ref{thermsusc}) by putting $m_\eta = m/N_f$ and $ \eta = 1/N_f$,
\be \chi(z_1,z_2)&=&-\frac{1}{16 \pi^2}
\left(\frac{\Sigma}{F^2}\right)^2 \Bigg[ \ln \frac{z_1+z_2}{2 \mu}
+\frac{z_1^2+z_2^2+2 m^2-2 (z_1+z_2) m }{N_f^2 (z_1-z_2)^2}
\nonumber \\ &+& \frac{2}{N_f^2(z_1-z_2)^3}\Bigg\{ \Big ( z_1^2
(z_1-3 z_2)+4 z_1 z_2 m-(z_1+z_2) m^2 \Big)  \ln \frac{ z_1}{\mu}
- z_1 \leftrightarrow z_2  \Bigg\}   \Bigg]. \nonumber \\
\label{thermoduqchi} \ee
 For $z_1\to z_2 =z$ the apparent singularities cancel, and in this limit
the expression (\ref{thermoduqchi}) simplifies to \be
\chi(z)=\frac{\Sigma^2 }{16 \pi^2  N_f^2 F^4} \Big( \frac{m^2}{3
z^2} + \frac{4 m}{3 z}-(N_f^2+2) \ln \frac{z}{\mu}  \Big), \ee or,
in the chiral limit, \be \chi(z)=-\frac{\Sigma^2 (N^2_f+2)}{16
\pi^2  N_f^2 F^4} \ln \frac{z}{\mu}. \ee Finally, in the case
$z_1=z_2=m$, we recover the result derived in \cite{SmilVerba}
within ChPT.

\subsection{Two-point correlation function}

In this section, we calculate the two-point spectral correlation
function from the discontinuities  of the disconnected scalar
susceptibility (see (\ref{chidisc})) obtained perturbatively
within pqChPT in the previous section. Therefore, our results for
the two-point function are only valid within the domain of
validity of perturbative pqChPT, \be 1/\Sigma L^4 &\ll& \lambda_i
\ll  \Lambda_{\rm QCD}  \; \; \; i=1,2,
\nonumber \\
 |\lambda_1 -\lambda_2| &\gg& \lambda_{\rm min} \sim 1/\Sigma L^4.
\label{range} \ee The second condition arises from the
susceptibilities $\chi(\pm i\lambda_1 +\epsilon, \mp i\lambda_2 +
\epsilon)$ that contribute to the two-point correlation function
(see eq. (\ref{chidisc})). In that case,  we have  zero-momentum
modes with mass $\sim |\lambda_1-\lambda_2|$ and a perturbative
evaluation of the susceptibility is only possible for  $\Sigma
V|\lambda_1 -\lambda_2| \gg 1$. For eigenvalues inside the domain
(\ref{range})  satisfying  the additional  condition 
$\lambda_i \ll  F^2/\Sigma L^2$, that is the ergodic domain, only the zero
momentum modes contribute to the scalar susceptibility which is
then given by (\ref{sus01}). For the connected two-point
correlation function we then find
\begin{eqnarray}
\label{CorrelErgodicCHPT} \rho_c(\lambda_1,\lambda_2)= -\frac{1}{2
\pi^2} \left[ \frac1{(\lambda_1-\lambda_2)^2 } +
\frac1{(\lambda_1+\lambda_2)^2 }
  \right] + \cdots.
\end{eqnarray}
This result is also obtained from a perturbative expansion of the
chRMT result, that is the expression (\ref{CorrelQuantum}) derived
from the zero-momentum sector of the pqChPT partition function in
Section~3.2. It does not depend on any of the low-energy coupling
constants that appear in the effective Lagrangian (\ref{Leff}) and
is thus the same, independent of the number of flavors and the
quark masses, provided that chiral symmetry is broken
spontaneously.

The perturbative result for the two-point spectral correlation
function is obtained from the discontinuity of (\ref{suscfirst}).
At finite volume, in   the domain (\ref{range}), it is given by
\begin{eqnarray}
  \label{finiteVcorrel}
  \rho_c(\lambda_1,\lambda_2)&=&-\frac{1}{2 \pi^2}
\frac{1}{(\lambda_1-\lambda_2)^2}
-\frac{1}{2 \pi^2} \frac{1}{(\lambda_1+\lambda_2)^2} \nonumber \\
&&+\frac{\Sigma^2}{2 \pi^2 F^4} \sum_{\vec{p} \neq 0} \Big\{
\frac{p^4-\frac{\Sigma^2}{F^4} (\lambda_1-\lambda_2)^2}
{(p^4+\frac{\Sigma^2}{F^4} (\lambda_1-\lambda_2)^2)^2}
+\frac{p^4-\frac{\Sigma^2}{F^4} (\lambda_1+\lambda_2)^2}
{(p^4+\frac{\Sigma^2}{F^4} (\lambda_1+\lambda_2)^2)^2}  \Big \}
\nonumber \\
&&  +  \frac{\Sigma^2}{2\pi^2 F^4}\sum_{\vec{p}\ne 0}4 (\eta p^2
+2\Sigma m_\eta/F^2)^2 \frac{p^4-\frac{\Sigma^2}{F^4} \lambda_1^2}
{(p^4+\frac{\Sigma^2}{F^4} \lambda_1^2)^2}
\frac{p^4-\frac{\Sigma^2}{F^4} \lambda_2^2}
{(p^4+\frac{\Sigma^2}{F^4} \lambda_2^2)^2}
  \Big\}.
\end{eqnarray}
In the ergodic domain, only the contribution of the first two
terms has to be taken into account. In the diffusive domain, the
thermodynamic limit of (\ref{finiteVcorrel}) is given by \be
\label{CorrelDiff} \rho_c(\lambda_1,\lambda_2)&=&-\frac{\Sigma^2
V}{64 \pi^4 F^4} \Bigg[ \ln \frac{(\lambda_1^2-\lambda_2^2)^2}{16
\mu^4} +8 \frac{(\lambda_1^2+\lambda_2^2) (\eta^2 (\lambda_1^2
+\lambda_2^2)-2 m_\eta^2)} {(\lambda_1^2-\lambda_2^2)^2}
\nonumber \\
&& -16 \pi \frac{\eta m_\eta  |\lambda_1| |\lambda_2|}
{(|\lambda_2|+|\lambda_1|)^3} +4
\frac{1}{(\lambda_1^2-\lambda_2^2)^3} \Bigg\{ m_\eta^2
(\lambda_1^4+6 \lambda_1^2 \lambda_2^2
+\lambda_2^4) \ln \frac{\lambda_1^2}{\lambda_2^2} \nonumber \\
&& +\eta^2 \lambda_1^2 [\lambda_1^4-6 \lambda_1^2 \lambda_2^2 -3
\lambda_2^4] \ln \frac{\lambda_1^2}{\mu^2} -\eta^2 \lambda_2^2
[\lambda_2^4-6 \lambda_2^2 \lambda_1^2 -3 \lambda_1^4] \ln
\frac{\lambda_2^2}{\mu^2} \Bigg\}   \Bigg]. \ee This result can
also be obtained directly from the thermodynamic limit of the
scalar susceptibility in the diffusive domain (\ref{thermsusc}).
The two-point correlation function is even in both $\lambda_1$ and
$\lambda_2$ but is not translational invariant. These properties
originate from the pairing of the Dirac eigenvalues. For massless
quarks (or topological susceptibility equal to zero in the
quenched case) the expression (\ref{CorrelDiff}) simplifies to
\begin{eqnarray}
\rho_c(\lambda_1,\lambda_2)&=&-\frac{\Sigma^2 V}{64 \pi^4 F^4}
\Bigg[ \ln \frac{(\lambda_1^2-\lambda_2^2)^2}{16 \mu^4} +\frac{4
\eta^2}{(\lambda_1^2-\lambda_2^2)^2} \Bigg\{ 2
(\lambda_1^2+\lambda_2^2)^2 \label{cormassless} \\
&& +\frac{1}{\lambda_1^2-\lambda_2^2} \Big( \lambda_1^2
[\lambda_1^4-6 \lambda_1^2 \lambda_2^2 -3 \lambda_2^4] \ln
\frac{\lambda_1^2}{\mu^2} - \lambda_2^2 [\lambda_2^4-6 \lambda_1^2
\lambda_2^2 -3 \lambda_1^4] \ln \frac{\lambda_2^2}{\mu^2} \Bigg)
\Bigg\}  \Bigg], \nonumber
\end{eqnarray}
where $\eta=1/N_f$ in the QCD case, and $\eta =\alpha$ in the
quenched case. Notice that in the limit $\lambda_1 \to \lambda_2$
the terms proportional to $\eta^2$ are regular. For
$|\lambda_1-\lambda_2| \ll \lambda_{1,2}$ the infrared singular
part of the correlation function  (\ref{cormassless}) simplifies
to
\begin{eqnarray}
  \rho_c(\lambda_1,\lambda_2)=-\frac{\Sigma^2}{32 \pi^4 F^4}
\log \frac{|\lambda_1-\lambda_2|}{\mu}.
\end{eqnarray}
This result, derived for $m_\eta = 0$, does not depend on the
parameter $\eta$. It is therefore valid for QCD with any number of
flavors of massless quarks, even zero. It cannot be obtained from
chRMT. Beyond the energy scale for which the kinetic term in the
chiral Lagrangian cannot be neglected, also known as the Thouless
energy, pqChPT differs from chRMT. This  was observed earlier in
the analysis of the spectral density in \cite{Vplb,OTV} and will
be discussed in greater detail in the next section.

\section{Number Variance}

In the study of disordered systems, a frequently used measure of
the spectral correlations is the number variance of the
eigenvalues. It is defined as the variance of the number of
eigenvalues in an interval that contains $n$ eigenvalues on
average. If the actual number of eigenvalues in each such interval
for the $i$-th member of the ensemble is given by  $n_i$, the
$p$-th moment of the number of eigenvalues is given by the
ensemble average $\langle n_i^p \rangle$. With the number variance
denoted by $\Sigma^2(n)$ we thus have \be \label{NV}
   n &=& \langle n_i \rangle,\nonumber\\
  \Sigma^2(n)&=&\langle n^2_i \rangle  - \langle n_i \rangle^2.
\ee

If we denote the eigenvalues for the $i$-th member of the ensemble
by $\lambda_k^{(i)}$ the corresponding spectral density is given
by \be \rho_i(\lambda) = \sum_k \delta(\lambda -\lambda_k^{(i)}).
\ee The number of eigenvalues inside the interval $[a,b]$ is equal
to \be n_i = \int_a^b d\lambda \rho_i(\lambda). \ee The average
number of eigenvalues inside this interval is thus given by \be n
= \int_a^b d\lambda \langle \rho_i(\lambda)\rangle, \ee and the
number variance can be written as \be \Sigma^2(n)
=\int_a^b\int_a^b d\lambda_1 d\lambda_2
\rho_c(\lambda_1,\lambda_2), \ee with the connected two-point
correlation function given by \be \rho_c(\lambda_1,\lambda_2) =
\langle  \rho_i(\lambda_1) \rho_i(\lambda_2)\rangle - \langle
\rho_i(\lambda_1)\rangle \langle \rho_i(\lambda_2)\rangle. \ee If
the average spectral density is not constant on the interval
$[a,b]$ there will be a contribution to the number variance due to
the variation of the average spectral density. We will always
assume that this contribution has been eliminated by a procedure
called unfolding \cite{Mehta}.

The two-point correlation function can be decomposed as \be
\langle  \rho_i(\lambda) \rho_i(\lambda')\rangle = \delta(\lambda
-\lambda') \sum_k \langle \delta(\lambda -\lambda_k)\rangle +
\sum_{k\ne l} \langle \delta(\lambda -\lambda_k)\delta(\lambda'
-\lambda_l) \rangle . \label{two+self} \ee The connected two-point
correlation function can be written as (see (\ref{rho2}))       
\be \rho_c(\lambda,
\lambda') = \delta(\lambda-\lambda') \rho(\lambda) + R(\lambda,
\lambda'), \ee where the two-point cluster function $R(\lambda,
\lambda')$ contains only correlations of different eigenvalues and
is regular for $\lambda \to \lambda'$.

For a finite total number of eigenvalues, we obtain by integration
of the connected two-point correlation function the sum rule \be
\int d\lambda_1 \rho_c(\lambda_1, \lambda_2) = 0. \label{sum-rule}
\ee The contribution due to the self-correlations is canceled by
the contribution from the two-point cluster function. The
coefficient of the linear term in the asymptotic expansion of the
number variance is given by \be \frac {d \Sigma^2(n)}{dn} = 2\frac
\pi{\Sigma V} \int_0^{\pi n/\Sigma V} d\lambda \rho_c(\frac{\pi
n}{\Sigma V}, \lambda). \ee where we have used that
$\rho_c(\lambda_1, \lambda_2) = \rho_c(\lambda_2, \lambda_1)$.
This quantity is also known as the spectral compressibility. We
observe that for large $n$ the coefficient of the linear term
vanishes provided that the correlation function approaches zero
faster than  $1/\lambda, \, 1/\lambda'$, i.e. when the integration
in the sum-rule (\ref{sum-rule}) is convergent without imposing a
cutoff on the total number of eigenvalues. If this is the case the
correct asymptotic result for the number variance is              
obtained from the asymptotic result of the two-point cluster
function provided it is regularized such that the sum-rule  
(\ref{sum-rule}) is satisfied.

In the previous sections, the two-point correlation function was
derived under the assumption that $\lambda_1, \lambda_2 \ll
\Lambda_{\rm QCD}$. On this scale, the variations of the average
spectral density \cite{OTV} can be neglected. The average number
of eigenvalues in the interval $[0,\Delta]$ is given by
\begin{eqnarray}
n   = \int_0^\Delta d \lambda \rho(\lambda).
\end{eqnarray}
For $m_\eta =0$ the spectral density for $ \lambda \ll
\Lambda_{\rm QCD}$ is given by the Banks-Casher relation
\begin{eqnarray}
  \label{density}
  \rho(\lambda)=\frac{\Sigma V}\pi+\cdots.
\end{eqnarray}
The average number of eigenvalues in the interval $[0,\Delta]$, of
eigenvalues is thus given by
\begin{eqnarray}
  n  = \frac{\Sigma V \Delta}\pi+\cdots,
\end{eqnarray}
and the number variance can be written as
\begin{eqnarray}
  \label{NumbVarUnf}
  \Sigma^2(n)&=&
\int_0^{\pi n/\Sigma V} d
  \lambda_1  \int_0^{\pi n/\Sigma V} d \lambda_2 \;
       \rho_c(\lambda_1,\lambda_2) \nonumber \\
&=& n + \int_0^{\pi n/\Sigma V} d
  \lambda_1  \int_0^{\pi n/\Sigma V} d \lambda_2 \;
       R(\lambda_1,\lambda_2).
\end{eqnarray}

Perturbative calculations of the spectral two-point function are
only valid for $|\lambda_1 - \lambda_2|\gg \lambda_{\rm min}$ and
do not include the term due to self-correlations. However, they
are included in the nonperturbative result (\ref{suscFinal}) valid
in the ergodic domain and in the quantum domain. Therefore the
number variance in the quantum domain and in the ergodic domain is
obtained by simply inserting the non-perturbative result
(\ref{CorrelQuantum}) into (\ref{NumbVarUnf}). We are not aware of
any analytic expression for these integrals. However both in the
case of small $n$ and in the case of large $n$, one can derive
concise analytic expressions for the number variance. Only the 
self-correlations in (\ref{CorrelQuantum}) contribute to the   
small-$n$ limit of the number variance,                        
\be  
\Sigma^2(n) = n + O(n^2). 
\ee

As discussed above the asymptotic large-$n$ result of the number
variance can only be obtained from the asymptotic result of the
two-point correlation function after it has been regularized such
that the sum-rule (\ref{sum-rule}) is satisfied. In the ergodic
domain this is achieved by
\begin{eqnarray}
\label{CorrelErgodicCHPTa} \rho_c(\lambda_1,\lambda_2)=
\rho(\lambda)[\delta(\lambda-\lambda') + \delta(\lambda+\lambda')]
-\frac{1}{2 \pi^2} \left[ \frac1{(\lambda_1-\lambda_2)^2 + a^2} +
\frac1{(\lambda_1+\lambda_2)^2 +a^2}
  \right] + \cdots. \nonumber\\                             
\end{eqnarray}
where $a$ is determined by the sum rule (\ref{sum-rule}), \be a =
\frac 1{2\pi \rho(\lambda)}. \ee In the ergodic domain, we have
that $\lambda \ll 1/L^2 \Lambda_{\rm QCD}$ so that $\rho(\lambda)$
is well approximated by $\rho(0)$. Then $\rho(0)$ drops out of the
expression for the number variance, and at leading order we find
the familiar logarithmic dependence known from Random Matrix
Theory, \be \Sigma^2(n) = \frac 1{2\pi^2} \log n + \cdots. \ee It
coincides with the result derived from  chRMT. Notice that in the
bulk of the spectrum the coefficient of $\log n$ is a factor of 2
larger.

\begin{center}
\begin{figure}[ht!]
\hspace*{1cm} \epsfig{file=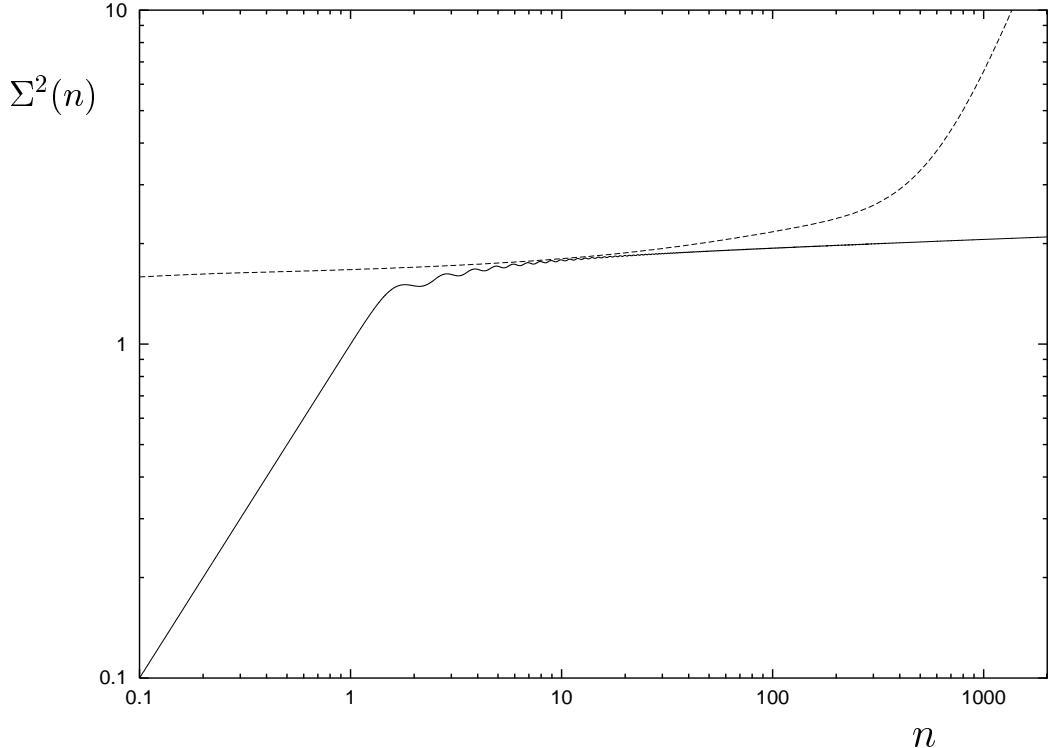, width=14.3cm, height=10cm}
\caption[]{\small The number variance for levels below 
$n_{\rm QCD}$ computed from pqChPT for $N_f=3$ massless quarks. The 
solid curve represents the result from  
the zero-mode sector of the partition function, or
chRMT, and the dashed curve shows the perturbative result. The volume 
of the box is $(20 \; {\rm fm})^4$, the Thouless scale is
$n_c\approx27.5$ and $n_{\rm QCD}\approx 10^5$. Notice how late
the asymptotic $\log n$-behavior is reached, and how early the
non-zero momentum corrections are visible. A strikingly similar
figure has been obtained for a disordered metal in
\cite{BraunMontambaux}.}
\end{figure}
\end{center}

At finite volume, for eigenvalues much larger than the smallest
eigenvalues but in the domain of chiral perturbation theory (i.e.
$n_{\rm QCD} \gg n \gg 1$), the number variance can be calculated
from the expression for the two-point correlation function that
includes the nonzero-momentum modes (\ref{finiteVcorrel}). We find
\begin{eqnarray}
  \Sigma^2(n)&=&\frac{1}{2 \pi^2} \log n
+\frac{1}{4 \pi^2} \sum_{\vec{k}\neq0} \log \frac{k^4+
\left(\frac{n}{2 \pi^2 n_c}\right)^2}{k^4} \nonumber \\
&&+8 \pi^2 \left(\frac{n}{2 \pi^2n_c}\right)^2
\sum_{\vec{k}\neq0} \left( \frac{\frac {\eta k^2}{4\pi^2} +
\frac{2 \Sigma m_\eta}{F^2}}{k^4+ \left(\frac{n}{2 \pi^2n_c}
\right)^2} \right)^2, \label{finiteVNumbVar}
\end{eqnarray}
where $n_c=m_c/\lambda_{\rm min}=F^2 L^2/\pi$ is the dimensionless 
Thouless scale. We have used that $p_\mu =2 \pi  k_\mu /L$ with
$k_\mu$  a four-dimensional hypercubic lattice with unit cell of
length one.

In the diffusive domain, the thermodynamic limit of
(\ref{finiteVNumbVar}) is given by
\begin{eqnarray}
  \Sigma^2(n)=- \frac{1+2 \eta^2}{16 \pi^4} \left(\frac{n}{n_c} \right)^2
\log \frac{n}{n_{\rm QCD}},
\end{eqnarray}
where $n_{\rm QCD} \equiv \mu/ \lambda_{\rm min}\sim \Lambda_{\rm
QCD} /\lambda_{\rm min}$. Therefore, for QCD with $N_f$ flavors of 
massless quarks one finds that in the diffusive domain $n_c \ll n
\ll n_{\rm QCD}$,  the number variance of the Dirac spectrum is 
given by
\begin{eqnarray}
  \Sigma^2(n)=- \frac{N_f^2+2}{16 \pi^4 N_f^2} \left(\frac{n}{n_c} \right)^2
\log \frac{n}{n_{\rm QCD}}.
\end{eqnarray}
The number variance computed from pqChPT for eigenvalues below       
$\Lambda_{\rm QCD}$ is shown in Fig.~2.                              

~Finally in the ballistic domain, for $n \gg n_{\rm QCD}$ the
eigenvalues of the Dirac operator in QCD are correlated as those
for a free Dirac operator. The $n^2 \log n$ behavior                   
saturates as $n$ for $n \gg n_{\rm QCD}$ where the interactions        
become weak because of asymptotic freedom.                             
The spectrum thus obeys Poisson
statistics, and the number variance is given by
\begin{eqnarray}
\Sigma^2(n)=n.
\end{eqnarray}

\section{Conclusions}

We have analyzed the fluctuations of the eigenvalues of the QCD
Dirac operator by means of the partially quenched chiral
susceptibility. The two-point spectral correlation function is obtained 
from its discontinuity across the imaginary axis.
The variance, 
$\Sigma^2(n)$, of the number of eigenvalues in an interval
containing $n$ eigenvalues on average is obtained by integrating
the spectral two-point correlation function. The generating
function for the scalar susceptibility is given by the QCD
partition function with two additional fermionic quarks and two
additional bosonic ghost quarks with a mass equal to the spectral
mass (also known as valence quark mass) that enters in the
partially quenched chiral susceptibility. For spectral quark
masses well below $\Lambda_{\rm QCD}$ the low-energy limit of this
theory is completely determined by its global symmetries.

Based on this partition function we have distinguished three
important scales in the Dirac spectrum, the smallest nonzero
eigenvalue, $\lambda_{\rm min}=\pi/\Sigma V$, the spectral mass
for which the Compton wavelength of the corresponding Goldstone
boson is equal to the size of the box, $m_c=F^2/\Sigma L^2$, and
$\Lambda_{\rm QCD}$. The analogues of these scales are well-known in 
mesoscopic physics where they separate the quantum domain, the
ergodic domain, the diffusive domain and the ballistic domain,
respectively.

For spectral quark masses well below $m_c$,  in the ergodic domain, 
the kinetic term in
the partially quenched chiral Lagrangian decouples from the zero
momentum part, and only the latter part has to be taken into
account. In this domain the partition function only depends on 
the number of flavors, the topological charge and the           
combination $mV\Sigma$ or $zV\Sigma$. Nontrivial results are obtained 
by keeping this combination fixed in the thermodynamic limit.
A perturbative expansion of the zero momentum part of  
the partition function is possible for spectral quark masses that are 
much larger than $\lambda_{\rm min}$. For smaller spectral 
quark masses we have                            
calculated exactly the super-integrals that appear in the 
partition function for the sector of topological charge $\nu$. Our
results for the scalar susceptibility  and the corresponding
spectral two-point correlation function in this domain are in
complete agreement with chiral Random Matrix Theory. This result,
together with earlier work on the microscopic spectral density,
provides strong evidence that also all higher order correlation
functions that can be derived from the low-energy limit of the QCD
partition function coincide with the results from chiral Random
Matrix Theory. The number variance in this domain is given by
$\log n/2 \pi^2$, showing that the fluctuations of the eigenvalues
are strongly suppressed.

For spectral masses in the diffusive domain, $m_c\ll z \ll
\Lambda_{\rm QCD}$, when nonzero momentum modes have to be taken into 
account, a perturbative evaluation of the partition is justified. 
In this limit for $\lambda \ll \Lambda_{\rm QCD}$ we have   
calculated the scalar susceptibility to one loop in        
chiral perturbation theory. We have found that in the case of
massive quarks (including the quenched case) that the scalar
susceptibility shows  a quadratic infrared divergence. At finite
volume this divergence is regulated by nonzero momentum modes, and
thus only appears for valence quark masses beyond the Thouless
energy. We expect that this  divergence will show up  much more 
prominently 
in lattice QCD simulations than its analogue of  quenched chiral  
logarithms. 

The prediction for the behavior of the two-point correlation
function in the diffusive regime is
$\rho_c(\lambda_1-\lambda_2)\sim
1/(\lambda_1-\lambda_2)^{(d-4)/2}$ \cite{altshuler-shklovskii}. In
agreement with this result we have found a logarithmic dependence
of $\rho_c(\lambda_1-\lambda_2)$ on $|\lambda_1 -\lambda_2|$. The
corresponding number variance is given by $\Sigma^2(n) \sim n^2
\log n$. The proportionality constant given by  
$(N_f^2 +2)/(16\pi^2 N_f^2 F^2 L^2)$ provides us with an alternative
method to determine the pion decay constant.

The behavior of the number variance is very different in the
various domains we have defined. Deep in the quantum domain, for
small $n$, the interval contains randomly either zero or one level
and $\Sigma^2(n)=n$. When the first level is reached, the number
variance almost stops increasing. It has a $\log n$ behavior up to
the Thouless scale $n_c=1/F^2 L^2$. Therefore the spectrum is
quite stiff in this domain. Around the Thouless scale, the number
variance increases much faster again. Well above $n_c$ but below
$n_{\rm QCD}$, the level number corresponding to $\Lambda_{\rm
QCD}$, the number variance grows like $n^2 \log n$. Finally, for
level number much larger than $n_{\rm QCD}$, the Dirac spectrum is
in essence a free particle spectrum because of asymptotic freedom.
The levels are therefore randomly distributed and the number
variance is again given by $\Sigma^2(n)=n$. This behavior is
typical for a disordered metallic phase in condensed matter 
physics.

In conclusion, we have shown that the fluctuations of the QCD
Dirac eigenvalues below $\Lambda_{\rm QCD}$ in the phase of broken
chiral symmetry are completely determined by the symmetries of the
QCD partition function.

\vskip 1.5cm \noindent {\bf Acknowledgements} \vskip 0.5cm

This work was partially supported by the US DOE grant
DE-FG-88ER40388. One of us (D.T.) was supported in part by
``Holderbank''-Stiftung and by Janggen-P\"ohn-Stiftung. D.
Dalmazi, P. Damgaard, J. Kogut, J. Osborn, K. Splittorff, and T. Wettig 
are acknowledged for useful discussions.

\end{document}